\documentclass[preprint]{elsarticle}
\usepackage{amsmath}
\usepackage{amsfonts}
\usepackage{amssymb,bm}
\usepackage{amsmath}
\usepackage{amsthm}
\usepackage{epsfig,epsf}
\usepackage{amsfonts,amscd,epsfig}
\usepackage{epstopdf}
\usepackage[english]{babel}
\usepackage{graphicx}
\usepackage{color}
\usepackage{epic}
\usepackage{eepic}
\journal{Physics Letters A}
\biboptions{square,comma}

\begin{document}
\begin{frontmatter}

\title{Fluxon interaction with the finite-size dipole impurity}
\author[is]{Ivan O. Starodub}
\ead{starodub@bitp.kiev.ua}

\author[yz]{Yaroslav Zolotaryuk\corref{cor1}}
\cortext[cor1]{Corresponding author}
\ead{yzolo@bitp.kiev.ua}

\address
{Bogolyubov Institute for Theoretical Physics, National Academy of
Sciences of Ukraine, Kyiv 03143, Ukraine}
\date{\today}

\begin{abstract}
Interaction of the fluxon with the finite size dipole impurity 
in the long Josephson junction is investigated. The impurity 
has polarity  and will be referred to as a
{\it dipole} impurity because it also has a direction and, consequently, 
changes its sign under the space inversion transform $x \to -x$.
Such a model is used to describe the inductively coupled to the 
Josephson transmission line qubit and the misaligned Abrikosov vortex 
that penetrates into the long Josephson junction. We derive the
approximate equations of motion for the fluxon center of mass and
its velocity. With the help of these equations we demonstrate that 
pinning and scattering of the fluxon on the impurity
differs significantly from the case of the point impurity which is modelled
by the derivative of the Dirac's $\delta$-function.
\end{abstract}

\begin{keyword}
josephson junctions \sep fluxon \sep soliton \sep impurity \sep
sine-Gordon equation
\end{keyword}

\end{frontmatter}

\section{Introduction}
Long Josephson junctions (LJJs) are of great interest for the
fundamental and applied physics \cite{barone82,u98pd}. 
Their practical applications range from astrophysics \cite{ks00sst} to 
quantum computation \cite{kwu02pssb,ars06prb,pkgku10prb,fssu13apl,fswbu14prl}. 
Among various wave phenomena in LJJs topological solitons occupy 
a special place. Fluxons or Josephson vortices are topological 
solitons of the sine-Gordon (SG) equation. A fluxon carries magnetic
flux quantum and is extremely robust because it is impossible to
destroy it with local deformations.

The idea of
the LJJ fluxon reading out the state of the JJ qubit
was first developed by Averin~\cite{ars06prb}. Later  
the problem of the LJJ fluxon interacting with the qubit
has been studied extensively both from the theoretical 
\cite{fssk-s07prb} and experimental \cite{fssu13apl,fswbu14prl} sides.
In theoretical studies the qubit is modelled as an $\delta'(x)$-impurity
in the LJJ where $\delta(x)$ is the Dirac's delta-function.
Such a choice of the impurity function was motivated mainly by
the simplicity of the mathematical operations with it. 

Another group of problems discussed in the literature is the 
fluxon scattering on the Abrikosov vortex.
First papers \cite{ag84jetpl} on the subject consider the Abrikosov vortex as
an impurity which possesses polarity and use the $\delta'(x)$
function to describe it in the respective equations of motion.
According to \cite{fg98prb} the Abrikosov vortex that penetrates in
the direction perpendicular to the junction plane should be described by the
spatially antisymmetric function similar to the $\delta'(x)$ function.
This Abrikosov vortex can be misaligned, thus, its size becomes finite 
and should be taken into account.
The junction with the dipole impurity was studied 
\cite{mu04prb} for the fluxon injection
mechanism in the annular JJ where the impurity described the pair of 
electrodes through which this injection process. In this case the
size of the current dipole was taken into account, but the problem
under consideration was rather different. The authors studied 
fluxon-antifluxon pair creation and fluxon scattering on the
dipole impurity with the antifluxon trapped there.

It should also be mentioned that the extensive research of the Schr\"odinger
equation with $\delta'(x)$ as a potential has 
revealed \cite{s86rmp,k96jmaa,cazeg03jpa,gm09umb,gh} that
this function is a non-trivial mathematical object that should be
treated with care. In particular, different approximating sequences 
of the $\delta'(x)$ can yield different results in the terms
of transmission, reflection and existence of the bound 
states \cite{gm09umb,zci06jpa,zz14ijmpb}.
Size effects play an important role in fluxon transmission through impurities
in one-dimensional \cite{kkc88pla,ddks12ejam} and 
two-dimensional \cite{m91pd,sz13ujp,sz14prb} Josephson junctions.

The aim of this paper can be easily formulated from the 
 above mentioned arguments. The question of
how the dipole impurity size influences the fluxon-impurity interaction in the
LJJ has not been studied yet and, consequently,
will be addressed in this paper.

The paper is organized as follows. In the next Section the model
is described. Section \ref{sec3} is devoted to the appoximate equations of
motion for the fluxon parameters. In Sec. \ref{sec4} the threshold
pinning current is computed as a function of the model parameters.
Delay time between pinning on the impurities with negative and
positive polarities is investigated in Sec. \ref{sec5}. 
Discussion and conclusions are given in the last Section.

\section{The model}
\label{sec2}
In this article the LJJ
with the dipole-like impurity is considered.
It is described by the perturbed sine-Gordon (SG) equation
\begin{equation}\label{1}
\phi_{tt}-\phi_{xx}+\sin \phi=-\alpha \phi_t+\gamma + f_{i}(x)~.
\end{equation}
This equation is written in the dimensionless form, where the
space variable is normalized to the Josephson length $\lambda_J$ 
and the time is normalized to the inverse Josephson plasma 
frequency $\omega_J^{-1}$. The function $\phi(x,t)$ is the
Josephson phase, which is the difference of the wave function
phases of the superconducting electrodes that constitute the junction.
The dimensionless parameters are:
$\alpha$ is the dissipation parameter, $\gamma$ is the
external bias current. The function $f_{i}(x)$ models the
dipole-like impurity
\begin{equation}\label{2}
f_{i}(x)= \mu \Delta_l' (x)=\sigma |\mu|\Delta_l' (x)~,~~
\sigma=\mbox{sign}(\mu).
\end{equation}
The parameter $\mu$ will be referred to as the impurity amplitude and
its sign $\sigma=\pm 1$ as its polarity.

Usually the Dirac's $\delta-$function and its derivatives are used 
as an approximation of the spatial inhomogeneities due to its 
simplicity. However, the inhomogeneities always have a finite size. 
Sometimes they are quite narrow and the approximation with the
$\delta'(x)$ function is valid, but often the size dependence is 
important. Therefore, the following approximation will be used
\begin{eqnarray}\label{imp}
&&\Delta_l'(x)=\left \{
\begin{array}{cc}
0,& |x|<\rho/2, \\
h,& -l-\rho/2<x<-\rho/2,\\
-h,& \rho/2<x<l+\rho/2,\\
0,& |x|>l+\rho/2\, .
\end{array} 
\right . =\\
&&=h\left\{\left[\theta\left(x+{\rho \over 2}+l\right)-
\theta\left(x+{\rho \over 2}\right)\right]
-\left[\theta\left(x-{\rho\over 2}\right)-
\theta\left(x-{\rho\over 2}-l\right)\right]\right\}~,
\nonumber
\end{eqnarray}
where $\theta(x)$ is the Heaviside step function. Here we assume 
that the total impurity length is $\rho+2l$ and it 
has an empty core with the length $\rho$. In the limit $l,\rho\to 0$
and $h=1/l^2 \to \infty$ one obtains $\Delta_l'(x) \to \delta'(x)$.

\section{Approximate equations of motion}
\label{sec3}

We apply the standard McLaughlin-Scott (MS) perturbation theory 
\cite{ms78pra} where the problem with the infinite-dimensional 
phase space is mapped
into the two-dimensional phase space with two dynamical variables. 
In the lowest order this theory states that
only the soliton parameters evolve in time as a response to the
perturbation, while the shape of the 
soliton remains unchanged. The SG soliton solution 
$\phi_0=4 \arctan \exp {[x-X(t)]/\sqrt{1-v^2(t)}}$ with its
parameters, the velocity $v$ and the center of mass coordinate $X$ 
being already the
functions of time, is substituted into the perturbed SG equation. 
As a result one obtains the system of two nonlinear 
first-order ODEs for $X$ and $v$. In the higher orders the shape
changes can be computed as well, including the plane-wave
radiation effects \cite{kkc88pla,m91pd,sz14prb}. 
This theory was designed for the integrable
systems, however, for the the center of
mass dynamics in the one-soliton case, it belongs to 
 the more general
family of the collective-coordinate methods that work for the 
non-integrable field models like $\phi^4$ \cite{gs75prd}. 
The application of the MS theory for the many-soliton problems
becomes problematic because of the significant difference in 
the soliton-soliton interaction in the integrable and non-integrable systems.

Thus, the perturbed SG equation transforms into 
the system of two nonlinear ODEs for the fluxon center of mass $X$ 
and its velocity $v$:
\begin{eqnarray}\label{5}
\dot X&=&v-\frac{1}{4}v\sqrt{1-v^2}\int_{-\infty}^{+\infty}
\frac{\Theta f_i(\Theta)}{\cosh{\Theta}}d\Theta\;, 
~~\Theta=\frac{x-X}{\sqrt{1-v^2}},\\
\label{6}
\dot v&=&\frac{\pi \gamma}{4} (1-v^2)^{3/2}-\alpha v(1-v^2)-
\frac{\sqrt{1-v^2}}{4}\int_{-\infty}^{+\infty}
\frac{f_i(\Theta)}{\cosh{\Theta}}\,d\Theta.
\end{eqnarray}
Here the dot corresponds to the time differentiation. 
After substituting Eq.~(\ref{imp}) into Eqs. (\ref{5})-(\ref{6}) and 
integrating over the region of impurity 
existence we obtain the dynamical equations for the fluxon
parameters:
\begin{eqnarray}\label{7}
&&{\dot v}=\frac{1}{4}\pi\gamma \left(1-v^2\right)^{3/2}-\alpha v
\left (1-v^2\right)-\frac{1}{2}h\mu\left(1-v^2\right)^{3/2}\Xi(X,v),\\
\label{8}
&&{\dot X}=v+\frac{1}{4} h\mu v\left(1-v^2\right)
\left \{\frac{2X~ \Xi(X,v)-\rho~ \Psi(X,v)}{\sqrt{1-v^2}} + \right .\\ 
\nonumber
&&+\frac{2 l}{\sqrt{1-v^2}}\arctan\left[\mbox{sech}
\left(\frac{X}{\sqrt{1-v^2}}\right)\sinh
\left(\frac{\rho/2+l}{\sqrt{1-v^2}}\right)\right] +\\ \nonumber 
&&+\Lambda \left(\frac{-\rho/2-l-X}{\sqrt{1-v^2}}\right)-
\Lambda \left(\frac{-\rho/2-X}{\sqrt{1-v^2}}\right)-
\Lambda \left(\frac{\rho/2-X}{\sqrt{1-v^2}}\right)+ \\ 
&&
\left .
+\Lambda \left(\frac{\rho/2+l-X}{\sqrt{1-v^2}}\right)\right\}~.
\nonumber
\end{eqnarray}
The auxiliary functions $\Xi(X,v)$, $\Psi(X,v)$ and $\Lambda(z)$ are 
introduced
for the sake of brevity and are given by the following expressions:
\begin{eqnarray}
\tan \Xi(X,v)&=& \frac{\left[\cosh \left(\frac{\rho}
{2\sqrt{1-v^2}}\right)-\cosh \left(\frac{l+\frac{\rho}{2}}{\sqrt{1-v^2}}
\right)\right]
 \sinh\left(\frac{X}{\sqrt{1-v^2}}\right)}{\sinh^2\left(\frac{X}
 {\sqrt{1-v^2}}\right)+\cosh \left(\frac{\rho}{2 \sqrt{1-v^2}}\right) 
 \cosh \left(\frac{l+\frac{\rho}{2}}{\sqrt{1-v^2}}\right)} ,\\
 \tan \Psi(X,v)&=& \frac{\left[\sinh \left(\frac{\rho}
{2\sqrt{1-v^2}}\right)-
\sinh \left(\frac{l+\frac{\rho}{2}}{\sqrt{1-v^2}}\right)\right] 
\cosh\left(\frac{X}{\sqrt{1-v^2}}\right)}{\cosh^2
\left(\frac{X}{\sqrt{1-v^2}}\right)+\sinh \left(\frac{\rho}{2 \sqrt{1-v^2}}
\right) \sinh \left(\frac{l+\frac{\rho}{2}}{\sqrt{1-v^2}}\right)} , \\
 \Lambda(z)&=&-i \left [\mbox{Li}_2 (i e^z)-\mbox{Li}_2 (-i e^z)\right ],
\end{eqnarray}
where $\mbox{Li}_2(x)$ is the polylogarithm function \cite{as84}
\footnote{
Here used the following expressions of polylogarithm function:
\begin{equation}
{Li}_1(x)=-ln(1-x),~~{Li}_{n+1}(x)=\int_0^x\frac{{Li}_n(t)}{t} dt~.
\end{equation}
}. There are two fixed points of this equation at which the fluxon 
can be found at rest ($\dot{X}=0,~v=0$):
\begin{eqnarray}
\label{12}
&&X_{\pm}=\sigma\mbox{arcsinh}\left[\frac{\sinh\left(\frac{\rho+l}{2}\right)
\sinh\left(l/2\right)}{\tan\left(\frac{\pi\gamma}{2|\mu| h}\right)}
\pm \right .\\ &&\left.
\nonumber
 \pm \frac{\sqrt{\sinh^2\left(\frac{\rho+l}{2}\right)
\sinh^2 {\frac{l}{2}}-\tan^2\left(\frac{\pi\gamma}{2\mu h}\right)
\cosh\left(\frac{\rho}{2}+l\right)\cosh\frac{\rho}{2}}}{\tan\left (
\frac{\pi\gamma}{2|\mu| h}\right)}\right],
\end{eqnarray}
where the inequality $|X_-|<|X_+|$ is always satisfied. 
If the bias is small enough and/or the impurity amplitude $\mu$ is large
enough, these two roots are real.
At this point it is hard to figure out which is a stable equilibrium and
which is unstable one and we will do it in the non-relativistic limit 
$v^2\ll 1$.
\par

It is easy to get the point impurity case if  $\rho=0, l\to 0, h\to 1/l^2$. 
Then the already known result of \cite{fssk-s07prb,ag84jetpl} is obtained:
\begin{eqnarray}\nonumber
\dot{v}&=&\frac{1}{4} \pi \gamma \left(1-v^2\right)^{3/2}-\alpha v
\left(1-v^2\right)+\\
&+&\frac{\mu}{4}\sqrt{1-v^2} \sinh\left(\frac{X}{\sqrt{1-v^2}}\right)
\text{sech}^2\left(\frac{X}{\sqrt{1-v^2}}\right)~,\\
\dot{X}&=&v+\frac{\mu}{4} v\left[{2}\cosh \left (
\frac{X}{\sqrt{1-v^2}}\right)-\frac{X}{\sqrt{1-v^2}}\sinh \left(
\frac{X}{\sqrt{1-v^2}}\right) \right]\times \\ 
&\times&\text{sech}^2\left(\frac{X}{\sqrt{1-v^2}}\right)~.
\nonumber
\end{eqnarray}

\begin{figure}
\centerline{\includegraphics[width=0.7\textwidth]{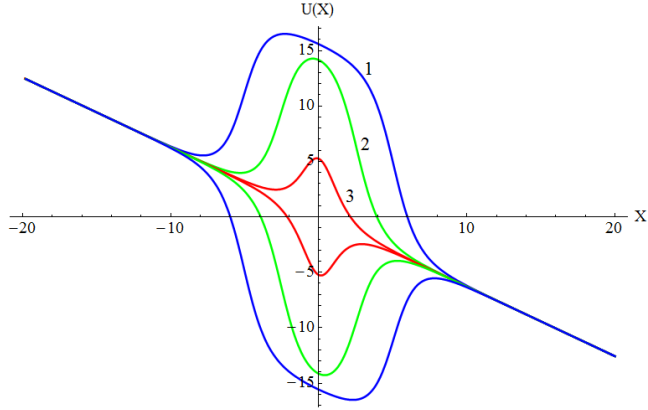}}
\vspace{2pt}
\caption{(Colour online) The potential $U(X)$ [Eq. (\ref{16})] 
created by the impurity with $\rho=10$ (blue, 1), $\rho=5$ (green, 2),
and $\rho=1$ (red, 3). Other parameters are: 
$\gamma=0.1,~ \mu=\pm 0.25$ and $l=0.1$.}
\label{fig1}
\end{figure}

Now we can find the potential created by the impurity (\ref{imp}). We
assume that the fluxon is in the non-relativistic regime ($v^2 \ll 1$).
In that case the system (\ref{7})-(\ref{8}) reduces to the
well-known Newtonian equation of motion 
\begin{eqnarray}\label{15}
&&8\ddot{X}+\alpha \dot{X} =-\frac{\partial U}{\partial X},\\
\label{16}
&& U(X)=-2\pi\gamma X+ U_0(X)=-2\pi\gamma X+4\mu h \int_{-\infty}^X \Xi(y,0)\,dy~, 
\end{eqnarray}
The exact shape of the potential $U_0(X)$ for the arbitrary barrier width $l$ 
and the distance between the impurity peaks $\rho$ cannot be 
obtained explicitly. Therefore we have used numerical integration in 
Eq. (\ref{16}). The resulting shape of $U(X)$ for different 
values of $l,~\rho$ and $\sigma$ is presented on Fig. \ref{fig1}. 
The tilt of the potential depends 
on the external current $\gamma$ while the potential term $U_0(X)$ in Eqs.
(\ref{15})-(\ref{16}) has the following general properties:
\begin{equation}
U_0(X)=U_0(-X)~,~~\lim_{|X|\to \infty }U_0(X)\to 0~,~~
U_0(X)=-U_0(X)~\mbox{if}~\sigma \to -\sigma~  .
\end{equation}
For $\sigma=1$ $U_0(X)$ is the barrier-like potential and for $\sigma=-1$
it is a potential well. 
This figure helps to understand the nature of the fixed points defined
by Eq. (\ref{12}).
For $\sigma=1$ both extrema lie on the negative half-axis 
with $X_{min}=X_+$, $X_{max}=X_-$, while for $\sigma<0$ they
lie on the positive half-axis: $X_{min}=X_-$, $X_{max}=X_+$. It is 
easy to see that the extrema of
potential $U(X)$ are symmetric with respect to the transform
$\sigma \to - \sigma$, $X_{max, min} \to -X_{min, max}$. In 
other words, when the polarity of the impurity changes, the maximum 
changes sign and becomes the minimum, or vice versa.

In the limiting case $\rho=0, l\to 0~ (h\to 1/l^2)$ the expression for 
the potential can be obtained explicitly:
\begin{equation}\label{potential}
U(X)=U_0(X)-2\pi\gamma X=2\sigma|\mu|~\text{sech}X-2\pi\gamma X~.
\end{equation}
The extrema of the potential (\ref{potential}) are given by the following
expression
\begin{eqnarray}\label{extremae}
&&X_{\pm}= \sigma \mbox{arcsinh}\left ( 
\frac{|\mu|}{2\pi\gamma} \pm \sqrt{\left(\frac{|\mu|}{2\pi\gamma} 
\right)^2-1} \right)\approx \\
&&\approx\sigma \left \{
\begin{array}{c}
\frac{\pi\gamma}{|\mu|}~(-)\\
\ln \frac{2|\mu|}{\pi\gamma}~(+)
\end{array}
\right ., ~~\gamma \ll|\mu|~ .
\nonumber
\end{eqnarray}

\section{Threshold current}
\label{sec4}

The current-voltage characteristic (CVC) for the dc-biased LJJ 
with an 
impurity usually has a hysteresis-like shape with two 
characteristic bias values:
the critical $\gamma_c$ current and the threshold current
$\gamma_{thr}<\gamma_c$ \cite{ms78pra}. The former depends solely on the
shape of the potential $U(X)$, and can be easily obtained as a condition
for $U(X)$ to have a local minimum. For the current problem it can 
be derived from positiveness of the expression under the square
root in Eq. (\ref{12}). 
The threshold current is the minimal current that is needed for 
a fluxon to pass the obstacle. In other words, at this value of the bias
current the LJJ CVC switches from the resistive to the superconducting branch. 
The threshold current depends not only on the impurity
parameters but on the dissipation as well. In this Section we will focus
on finding this current for different values of the impurity polarity 
$\sigma$.
The method, developed previously in \cite{kmn88jetp} will be 
used. The main idea of the method comes from the approximate integration
of the newtonian equation of motion for the fluxon center of 
mass (\ref{15})-(\ref{16}). The left and right sides of this 
equation can be multiplied by $\dot{X}$ and integrated with respect to
time along the interval $t \in [0, t_{stop}]$, where $t_{stop}$ is the
time moment when the fluxon stops. Assume that at $t=0$
the fluxon is launched at $X=-\infty$ with kinetic energy $E_k$.
If the bias is exactly the threshold
bias $\gamma_{thr}$, the fluxon will stop at the maximum of $U(X)$ at
$X_{max}=X(t_{stop})$. Then, the integration $\int_0^{t_{stop}}[\ldots]dt$
of the both sides of the equation of motion is equivalent to
the integration $\int_{-\infty}^{X_{max}}[\ldots] dX$ and the threshold
current should satisfy the following equation 
\begin{equation}\label{enerbal}
E_k+8\alpha \int_{-\infty}^{X_{max}}\left(\frac{\pi\gamma_{thr}}{4\alpha}
-\dot{X} \right) dX =U_0(X_{max}),
\end{equation}
where the fluxon kinetic energy $E_k$ and its velocity $v_\infty$  at 
the infinite distance away from the impurity $v_\infty$ (see Ref. 
\cite{ms78pra}) are
\begin{eqnarray}
&&{E}_k=8\left[(1-v_{\infty}^2)^{-1/2}-1\right]\simeq 4 v_\infty^2
+{\cal O} (v_\infty^4),~\\
&& v_{\infty}=\left[1+\left(\frac{4\alpha}{\pi\gamma}\right)^2\right]^{-1/2}
\simeq \frac{\pi\gamma}{4\alpha}+{\cal O} 
\left[\left(\frac{\pi\gamma}{4\alpha} \right)^2\right]~.
\end{eqnarray}
The equation (\ref{enerbal})  can be treated as the
energy balance equation $E_k+\Delta E=U(X_{max})$~.
For the case $\sigma=1$ it has the following physical interpretation:
the fluxon kinetic energy $E_k$ at the starting point
at $X\to -\infty$ is spent on surmounting the local barrier of
$U(X)$ and overcoming the dissipation effects (given by the term $\Delta E$).
For the case $\sigma=-1$ such an interpretation is not possible because
the fluxon receives additional acceleration before starting to climb
the barrier.

\subsection{Positive polarity case, $\sigma=1$}

First we consider the case of positive polarity ($\sigma=1$) for which 
the dissipative energy loss equals
\begin{equation}\label{23}
\Delta E=8\alpha\int_{-\infty}^{X_{max}}\left(v_{\infty}-
\dot{X}\right)dX~.
\end{equation}
We will consider the limit where $\gamma\ll\mu$, $\alpha\ll \mu$. Then, 
the kinetic energy of the fluxon will take the form: 
$E_k=4v_{\infty}^2\approx \left({\pi \gamma}/{2\alpha}\right)^2$. 
In the lowest order of approximation the threshold current can be 
calculated 
by equating the fluxon kinetic energy to the maximum height of the
potential barrier: $4v_{\infty}^2=2{\mu}$, 
$\gamma_{thr}^{(0)}=2\alpha\sqrt{2\mu}/\pi$.
On the other hand, the fluxon kinetic energy satisfies the following
$4\dot{X}^2=4v_{\infty}^2-U(X)$ (the dissipative effects are neglected
at this point). After substituting the potential  ${U}(X)$ from  
(\ref{potential}) we will find
\begin{equation}
\dot{X}^2=\pi\gamma X/2+
\frac{\mu}{2}(1-\text{sech}X)\approx \frac{\mu}{2}(1-\text{sech}X)~.
\end{equation}
Now, having the explicit form of $\dot X$ we can compute the 
energy losses due to dissipation:
\begin{eqnarray} \label{diss}
\Delta E &=&8\alpha\sqrt{\frac{\mu}{2}}\int_{-\infty}^{X_{max}}
\left[1-\sqrt{1-\text{sech}X}\right]dX = 
\nonumber \\ 
&=& 8\alpha\sqrt{\frac{\mu}{2}}\left [2~\mbox{arcsinh} 
\left(\cosh{X_{max}}\right )
+X_{max}-\ln{2} \right ] \simeq 
\nonumber \\
&\simeq & 8\alpha\sqrt{\frac{\mu}{2}} \left [2 \ln \left( 
\frac{\sqrt{2}+1}{\sqrt{2}}\right)+X_{max}+{\cal O}(X_{max}^2)
\right ]~,
\end{eqnarray}
where $X_{min}$, $X_{max}$ are the minima and maxima of $U(X)$.
In accordance with (\ref{extremae}) $X_{min}\simeq-\ln(2\mu/\pi\gamma)$, 
$X_{max}\simeq-\pi\gamma/\mu$. Substituting these expressions into 
(\ref{diss}) and using the expansions of terms with the inverse 
hyperbolic functions to the Taylor series will find the approximate 
expression for the dissipative energy loss:
\begin{equation}\label{deltae}
\Delta E\approx 4\alpha\sqrt{2\mu}\left
[\ln \left(\frac{1+\sqrt{2}}{\sqrt{2}}\right)-
\frac{\pi\gamma}{\mu}\right ]~.
\end{equation}
Here we have assumed $|X_{max}|\ll 1$ and, consequently, the terms
${\cal O}(X_{max}^2)$ have been dropped. 
Now we can substitute Eqs. (\ref{deltae}) and (\ref{potential}) into 
the energy balance equation (\ref{enerbal}). Details of the
calculation are given in \ref{app1}.
The final equation for threshold current reads:
\begin{equation} \label{threshold_mupos}
\gamma_{thr}=\frac{2\alpha}{\pi}\left[\sqrt{2\mu}-
4\alpha \ln \left(\frac{1+\sqrt{2}}{\sqrt{2}}\right)\right]~.
\end{equation}
If the terms ${\cal O}(\alpha^2)$ are ignored we obtain the
expression $\gamma_{thr}=\sqrt{8\mu} \alpha/{\pi}$ which can be
derived from purely kinematic approach as in \cite{ms78pra}.
This approach takes into account only the barrier height of the
potential $U(X)$ but not its shape. The ${\cal O}(\alpha^2)$ correction, 
on the contrary, accounts 
for the potential shape as it is different from the respective
correction for the microshort impurity  
\cite{kmn88jetp} where $U(X)\propto \cosh^{-2}X$.

\subsection{Negative polarity case $\sigma<0$}

Now consider the opposite case of negative impurity amplitude 
$\mu<0$ ($\sigma=-1$). For this case the extrema points 
are: $X_{max}\simeq\ln(2|\mu|/\pi\gamma)$, $X_{min}\simeq\pi\gamma/|\mu|$.
We seek the approximate solution of Eq. (\ref{enerbal}) under the same
assumption $\gamma\ll |\mu|$ and $\alpha \ll |\mu|$. The integral in the 
l.h.s of this equation is negative from the moment of the soliton 
launch till it passes the minimum of $U(X)$ because the fluxon moves
down the well, and, therefore, accelerates with respect to the 
equilibrium velocity $v_\infty$. 
After passing the minimum point it starts to slow down, and, eventually
stops at $X=X_{max}$ if $\gamma=\gamma_{thr}$. 
The largest contribution
to the integral in the l.h.s of this equation comes from the area 
$-1<X<1$ when the fluxon moves around the minimum and has the
largest velocity. Here we can assume that it moves according to the
equation of motion $4\dot{X}^2+U_0(X)\approx 0$, and, therefore
$X(t)\simeq \sqrt{|\mu|/2} \sqrt{\mbox{sech}X}$. We  substitute this
law of motion into the main equation (\ref{enerbal}). As a result
we get
\begin{eqnarray}\nonumber
&&8\alpha \int_{-\infty}^{X_{max}}(\dot {X}-v_\infty)dX\simeq \\
&& \label{27}
\simeq 8 \left \{\alpha\sqrt{|\mu|}~
{}_2F_1\left [{1\over 4},{1\over 2},{5\over 4},-\left(\frac{\pi\gamma}{2|\mu|}\right) ^2\right ]+\frac{\pi\gamma}{2}
\left (1+\ln \frac{\pi\gamma}{2|\mu|}\right ) \right \},
\end{eqnarray}
where ${}_2F_1(a,b,c,d)$ is the hypergeometric function \cite{as84}.
The last term in Eq. (\ref{enerbal}) equals 
\begin{equation}\label{28}
U_0(X_{max})=-4|\mu| \frac{\frac{\pi\gamma}{2|\mu|}}{
1+\left (\frac{\pi\gamma}{2\mu} \right)^2}\simeq -2\pi\gamma +
{\cal O}\left [\left (\frac{\pi\gamma}{2\mu}\right)^2 \right ]. 
\end{equation}
Finally, after some calculations which are presented in
\ref{app1} we obtain the final formula for the threshold current: 
\begin{equation}\label{29}
\gamma_{thr}=\frac{4}{\pi^{5/4}}\Gamma \left 
(\frac{1}{4} \right)|\mu|^{1/4}\alpha^{3/2}+12\frac{\alpha^2}{\pi}
\ln \left[ \left(\frac{2\Gamma\left 
(\frac{1}{4} \right)}{\pi^{1/4}}\right)^{2/3}
\frac{\alpha}{|\mu|^{1/2}}\right]~,
\end{equation}
where $\Gamma(x)$ is the gamma-function. 
This formula has a structure similar to the expression for 
$\gamma_{thr}$ for the microresistor \cite{kmn88jetp}. For example, 
it does not have the 
${\cal O}(\alpha)$ term and there is a ${\cal O}(\alpha^{3/2})$ term, 
which is the lowest order approximation.

\subsection{Threshold current for arbitrary values of $\rho$ and 
$l$.}
\label{arb_l}

\begin{figure}[htb]
\centerline{\includegraphics[width=1.15\textwidth]{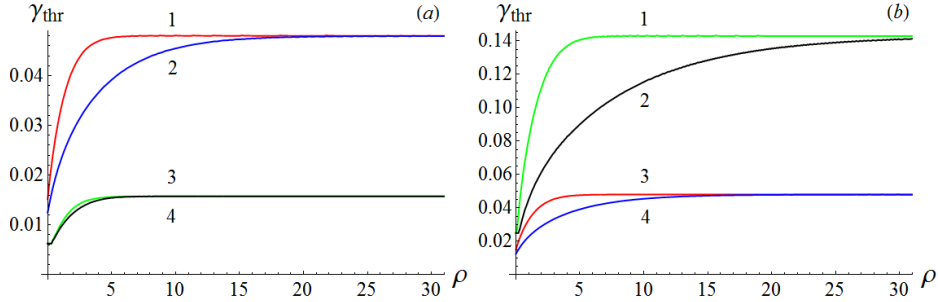}}
\vspace{2pt}
\caption{(Colour online) 
(a) Dependence of the threshold current on the impurity length 
$\rho$ for different values of $\mu$. For all curves 
$\alpha=0.1,~l=0.5$; $\mu=0.1$ (1,red), 
$\mu=-0.1$ (2, blue), $\mu=0.025$ (3, green), $\mu=-0.025$ (4, black).
(b)
Dependence of the threshold current on $\rho$ and $l$. 
For all curves $\alpha=0.1$; $\mu=0.1,~l=0.1$ (1, green),  
$\mu=-0.1,~l=0.1$ (2, black), 
$\mu=0.1,~l=0.5$ (3, red), $\mu=-0.1,~l=0.5$ (4, blue).}
\label{fig2}
\end{figure}

\begin{figure}
\centerline{\includegraphics[width=0.78\textwidth]{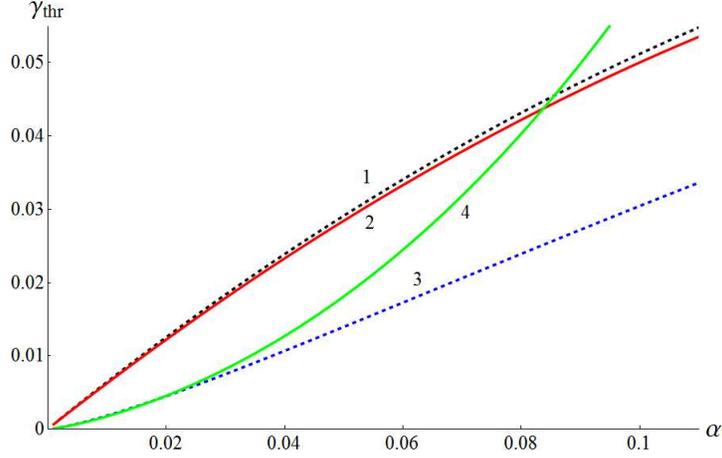}}
\vspace{0.5pt}
\caption{(Colour online) Dependence of the threshold current on $\alpha$. 
For all curves $\rho=0,~\mu=0.5,~l=0.001$. Dashed curves 1 (black) and 
3 (blue) represent the numerical results for $\sigma=1$ and $\sigma=-1$,
respectively.  Solid curves 2 (red) and 4 (green) show the analytical results
(\ref{threshold_mupos}) and (\ref{29}), for $\sigma=1$ and $\sigma=-1$
respectively.}
\label{fig3}
\end{figure}

\begin{figure}[htb]
\centerline{\includegraphics[width=1.1\textwidth]{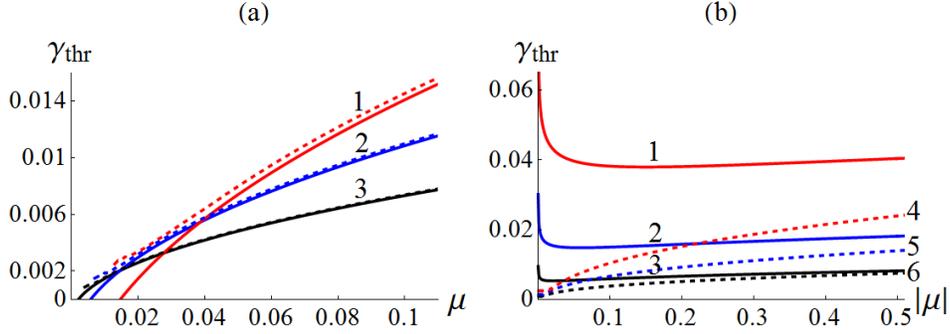}}
\vspace{0.5pt}
\caption{(Colour online) Dependence of the threshold current on the
impurity amplitude $\mu$ for $\rho=0$, $\sigma=1$ (a) and
$\sigma=-1$ (b). Numerically computed results are shown by
the dashed lines  and the solid lines 
correspond to Eqs. (\ref{threshold_mupos}) and (\ref{29}).
The dissipation values are $\alpha=0.08$ (curve 1, red), 
$\alpha=0.05$ (curve 2, blue) and $\alpha=0.03$ (curve 3, black).
The dashed curves in (b) correspond to $\alpha=0.08$ (curve 4, red),
$\alpha=0.05$ (curve 5, blue) and $\alpha=0.03$ (curve 6, black).
}
\label{fig3a}
\end{figure}

When the size parameters of the impurity, $l$ and $\rho$, are non-zero
it is not possible to obtain the threshold current analytically. Thus,
we have used numerical integration of the equations of motion 
(\ref{7}-\ref{8}). For this purpose we have
used the 4th order Runge-Kutta method. The polylogarithm 
function was computed with the help of the CHAPLIN library \cite{chap}.
In order to be able to get to the $\delta'(x)$ 
limit we have assumed $h=1/l^2$ throughout this and next Sections.
The numerically calculated threshold current values 
are presented in Figs. \ref{fig2}-\ref{fig3a}. 

Consider first Fig. \ref{fig2}, where the threshold current dependence
on the impurity length $\rho$ is presented. It appears that the
difference between $\gamma_{thr}$ for the impurities with different
polarities is not significant for small $\rho$'s but increases as
$\rho$ increases. Threshold current for $\sigma=1$ ($\mu>0$) is
always larger than for $\sigma=-1$. This can be understood from
the following argument. For $\sigma=-1$ the fluxon accelerates
all the way before the moment when it starts to climb the barrier.
For $\sigma=+1$ it slightly decelerates below the equilibrium
velocity $v_\infty$. Thus, in the $\sigma=-1$ case it needs less
energy to overcome the barrier of the same hight as compared with 
the $\sigma=1$ case. 
For larger values of $|\mu|$ the difference between $\gamma_{thr}$
for $\sigma=\pm 1$ is more pronounced, while it can be negligible 
if $|\mu|$ significantly decreases. The same occurs as $l$ is decreased
(see Fig. \ref{fig2}b). It should be noted that the impurity height 
is $h=1/l^2$, thus, decrease of $l$ in fact brings the effective 
increase of $|\mu|$.

%

It is interesting to discuss the limit of very long impurities,
when the distance between the extremal points of the impurity  
is large: $\rho\gg 1$. In this
case the threshold current should converge to the same value
for both polarities as can be observed in Fig. \ref{fig2}. 
Indeed, if $\sigma=1$, the fluxon approaches
the impurity with the velocity $v_\infty$ and it should be sufficient
enough to overcome the barrier $\Delta U=U(X_{min})-U(X_{max})$.
After it has climbed atop the barrier, the fluxon begins to 
slide down with the velocity which is almost the equilibrium
velocity $v_\infty$  because the slope of the barrier
is defined only by the bias value $\gamma$. 
If $\sigma=-1$ and the well is very wide, the fluxon falls into
the potential well and begins to move there again with the 
equilibrium velocity $v_\infty$. Thus, it approaches
the barrier with the equilibrium velocity and need to overcome the
barrier of the same height $\Delta U=U(X_{min})-U(X_{max})$. 
Therefore, the threshold current without the dissipative corrections
reads
\begin{equation}
\gamma_{thr}=\frac{2\alpha}{\pi}\sqrt{U(X_{min})-U(X_{max})}~.
\end{equation}

On the other hand, it is interesting to look at the dependence of 
threshold current from dissipation parameter which is presented 
in Fig. \ref{fig3}. For the positive polarity ($\mu>0$ or $\sigma=1$) the 
dependence is almost linear and has a good 
agreement of numerical and analytic results. 
For the opposite case of $\sigma=-1$  the agreement is quite 
good for small dissipation but as $\alpha$ increases up to $\alpha \sim 0.05$, 
the discrepancy
between Eq. (\ref{29}) and numerics becomes strong. At some point the
second term in 
(\ref{29}) becomes positive and the approximation breaks. Thus,  
the weakest condition for applicability of the 
expansion is $\alpha^2< \mu$.

%

Finally, we discuss the dependence of the threshold current on the 
impurity amplitude $\mu$. The respective dependences are given in 
Fig. \ref{fig3a}. For the positive polarity ($\sigma=1$) the threshold
current increases with $\mu$. This is quite natural, because the 
potential barrier hight increases. The analytical approximation 
(solid lines) works
well in comparison with the numerical data (dashed lines) 
for all $\mu$'s except the very small ones. The difference
between the numerical and analytical results decreases as the ratio
$\alpha^2/\mu$ decreases. 
In the limit
$\mu\to 0$ the analytical approximation (\ref{threshold_mupos}) 
breaks down because the second (negative) term becomes larger 
than the first one. This happens because this approximation is 
valid only under the condition $\alpha^2\ll \mu$. 
We have stopped the computation of $\gamma_{thr}$
for the very small values of $\mu$ because it required large
junction length and long calculation times. 

%

In the negative polarity case $\sigma=-1$ we observe
the non-physical divergence of the $\gamma_{thr}(|\mu|)$ dependence
[see Fig. \ref{fig3a}(b)].
It occurs again due to the fact that the analytical approximation is
not valid if $\mu\ll \alpha^2$. On the other hand, we observe the
convergence of the numerically computed (dashed) results with the
analytical approximation Eq. (\ref{29}) (solid lines) 
as $|\mu|$ increases. We remind
once again that this approximation works only if $\alpha^2\ll |\mu|$.
The fastest convergence is observed for the smallest dissipation value,
$\alpha=0.03$ (black curves 3 and 6). 
While for the $\sigma=+1$ case the threshold current increases as
$\gamma_{thr} \propto \mu^{1/2}$, for the negative polarity it seems 
to tend to some constant value. In reality it also increases, 
but as a $|\mu|^{1/4}$ function, which grows much slower.
Our calculations have been performed for $\rho=0$, however, as we know
from Fig. \ref{fig2}, the threshold current increases when $\rho$
 increases. Therefore, the $\gamma_{thr}(\mu)$ 
 dependence should be  modified accordingly for $\rho\neq 0$.

\section{Fluxon delay time on the impurity}
\label{sec5}

During its motion along the junction the fluxon meets the impurity and 
interacts with it, whereupon the fluxon velocity changes. After leaving
the impurity the fluxon velocity returns to the equilibrium value
 $v_\infty$. As it was
shown in the previous sections, the dynamics of the fluxon transmission
through the impurity with different $\sigma$ is quite different. In
particular, it takes different amount of time for the fluxon to
pass through the $\sigma=1$ and the $\sigma=-1$ impurities, respectively.
Thus, it is possible to define the delay time $\Delta t$ as a difference
between the time, necessary for the fluxon to pass some
fixed distance over the $\sigma=-1$ 
and $\sigma=+1$ impurities with all other parameters (including $\gamma$) fixed.
In \cite{ars06prb,fssk-s07prb} the process of the qubit read-out has been
proposed. The qubit is coupled to the Josephson transmission line and is
described theoretically as a dipole impurity. Its state is defined by
 the sign of $\sigma$. The delay time can be measured and is used to determine 
the state of the qubit.
 
\begin{figure}
\centerline{\includegraphics[width=0.8\textwidth]{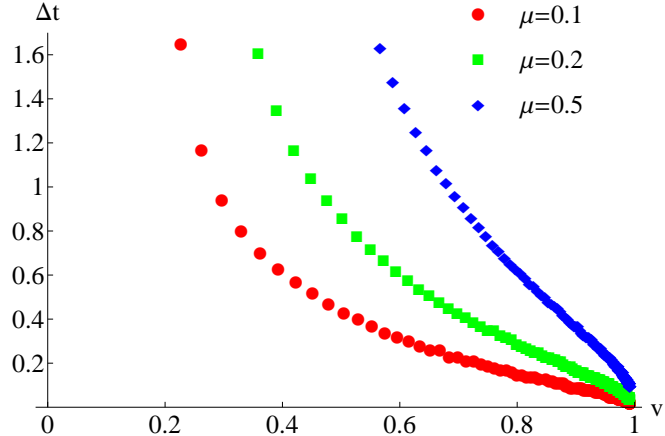}}
\vspace{2pt}
\caption{(Colour online) Fluxon delay time as a function of the fluxon 
velocity for the junction with $\alpha=0.1$ and impurity 
with $l=0.5,~\rho=0.1$ and different strength $|\mu|$ (shown in the
legend box).}
\label{fig4}
\end{figure}

\begin{figure}
\centerline{\includegraphics[width=0.8\textwidth]{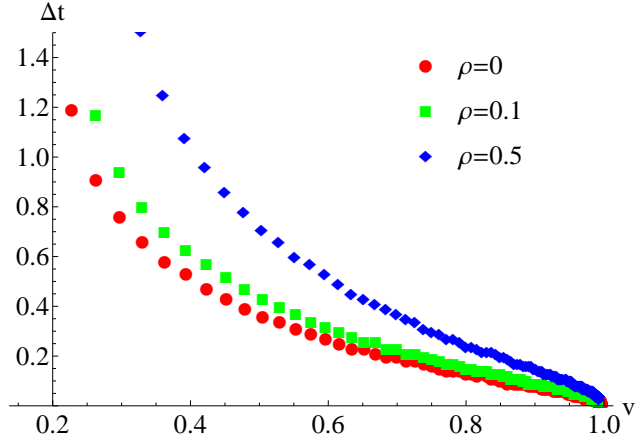}}
\vspace{2pt}
\caption{(Colour online) Fluxon delay time as a function of its velocity 
for the junction with $\alpha=0.1$ and impurity with $l=0.5,~|\mu|=0.1$ 
and different length $\rho$ (shown in the
legend box).}
\label{fig5}
\end{figure}

\begin{figure}
\centerline{\includegraphics[width=0.8\textwidth]{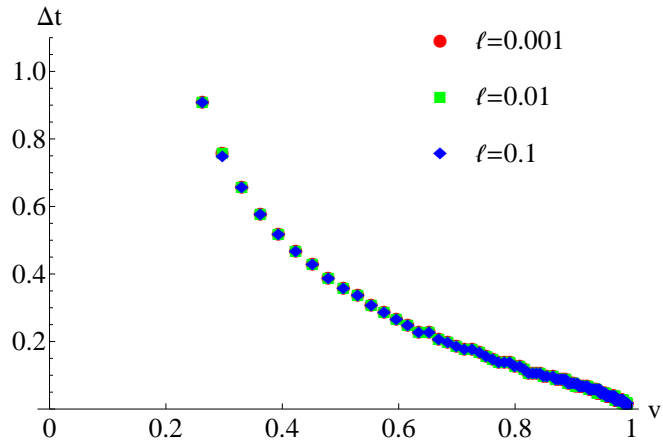}}
\vspace{2pt}
\caption{(Colour online) Fluxon delay time as a function of its velocity  
for the junction with $\alpha=0.1$ and impurity with $\rho=0,~|\mu|=0.1$ 
and different values of $l$ (shown in the
legend box).}
\label{fig6}
\end{figure}

In order to find the delay time we have performed numerical simulations 
of the fluxon equations  of motion (\ref{7})-(\ref{8}). The fluxon is 
launched at some distant point to the left from the impurity. Next, we 
measure numerically how much time has passed while fluxon got to the 
observation point on the right side from the impurity. Then 
the polarity of the impurity is changed and the same computation is repeated 
again. The difference between the obtained times is the fluxon delay  
time $\Delta t$. The absolute value of this delay time depends  on
 the parameters of the impurity and on the initial fluxon velocity 
 $v_\infty$, which itself depends on the external bias. 
The delay time as a function of the initial fluxon 
velocity is presented in Figs. \ref{fig4}-\ref{fig6} for the different
parameters of the impurity.

%

The main feature of all these three figures is the fact that the 
delay time decreases while the  fluxon 
velocity increases and tends to zero as $v\to 1$. This happens 
because the fluxon energy increases
in the relativistic limit as $\propto (1-v^2)^{-1/2}$, thus, the 
energy of the impurity-created potential barrier becomes negligible
comparing to the fluxon kinetic energy. As a result, there should be 
no difference in the fluxon transmission through the $\sigma=+1$
 and $\sigma=-1$ impurity at $v\to 1$. Also, if the bias current
 exceeds the critical current, there should be no pinning potential
 $U(X)$ at all.
 All the dependencies start from some certain finite
 value which is defined by the threshold current ${\gamma_{thr}}_{\sigma=+1}$. 
From Sec. \ref{sec4} we already know that 
${\gamma_{thr}}_{\sigma=+1}>{\gamma_{thr}}_{\sigma=-1}$
for all other parameter values.
Therefore, we have performed our computations for the bias 
$\gamma>{\gamma_{thr}}_{\sigma=+1}$. Calculations for the bias in
the interval
${\gamma_{thr}}_{\sigma=-1}<\gamma<{\gamma_{thr}}_{\sigma=+1}$
do not makes sense since $\Delta t$ would be infinity simply
because for $\sigma=1$ the fluxon would never arrive at the
measurement point. However, it does not mean that the experimental
read-out process is not possible for this range. Moreover, 
in this interval it will be the most efficient.

In Fig. \ref{fig4} the $\Delta t(v)$ dependence for
different values of the impurity amplitude $|\mu|$ is shown.
The shape parameters of the impurity, $l$ and $\rho$ 
[see Eq. (\ref{imp})] are fixed. The main feature of this
graph is that the delay time increases considerably (several times)
when the impurity amplitude $|\mu|$ increases. 
The similar dependence in 
Fig. \ref{fig5} manifests that $\Delta t(v)$ increases 
when the impurity length $\rho$ is 
increasing but the width of the barrier $l$ is fixed. 
We observe that the total impurity length
influences the delay time. 
If $\rho$ is increased, the delay time increases as well and this 
increase can be up to factor two. When the change of $l$ is concerned,
the delay time is much less sensitive to the change of this parameter.
In Fig. \ref{fig6} we observe that there is no visible change in
$\Delta t$ when $l$ is increased from $l=0.001$ to $l=0.1$. 
If one considers a quite moderate change of $\rho$ from $0$ to $0.1$ 
%
in Fig. \ref{fig5} (compare red and green markers), the increase
of $\Delta t$ is small but is still clearly visible. 
The very weak dependence on the parameter $l$ is explained by the fact that 
the height of the barrier equals $h=1/l^2$ and the impurity amplitude $\mu$
is multiplied by this coefficient [see Eqs. (\ref{7})-(\ref{8})]. 
Thus, the increase of the impurity length is compensated by the 
effective decrease of its amplitude. In fact, the parameter $l$ measures
the deviation from the $\delta'(x)$ approximation. We see that this 
approximation works fairly well. 
The parameter $\rho$, on contrary, is not connected to other 
system parameters
and controls the proper width of the potential barrier $U(X)$ that
is created by the qubit.

\section{Discussion and conclusions}
\label{sec6}
In this paper the fluxon interaction with the dipole-like impurity in the
long Josephson junction (LJJ) has been investigated. The problem arises
when studying the fluxon-Abrikosov vortex interaction 
\cite{ag84jetpl,fg98prb} in LJJ and
the qubit state read-out process \cite{ars06prb,fssu13apl,fssk-s07prb} 
in the Josephson transmission line
coupled to the qubit.  At variance with the
previous research we take into account the finite size of the 
impurity. The impurity is modelled by the piecewise function that
contains the well and barrier with the width
$l$ each that are separated from each other by the distance $\rho$. 
The impurity 
remains antisymmetric and reduces to the $\delta'(x)$ function in the
limit when the barrier(well) width becomes infinitely small. 

The main results of this research can be formulated as follows. 
The physical size of impurity influences considerably the fluxon-impurity
interaction. The most interesting result is the strong dependence of the
threshold pinning current on the impurity length, $\rho$.
The threshold current increases monotonically as $\rho$ increases, 
however, this is saturation growth which tends to some constant value
of $\gamma_{thr}$. The growth almost stops at some critical value
that  
equals several units of $\lambda_J$ (Josephson penetration depth) and 
depends on the impurity amplitude. There exist an optimal range
of $\rho$ for which the difference between 
${\gamma_{thr}}_{\sigma=-1}$ and ${\gamma_{thr}}_{\sigma=+1}$ is maximal. 
Thus, the
 range for the external bias where the 
 ${\gamma_{thr}}_{\sigma=-1}<\gamma<{\gamma_{thr}}_{\sigma=+1}$
can be increased considerably if $\rho$ and/or the impurity amplitude 
$\mu$ are chosen
appropriately. In this range the qubit read-out 
 process is the most efficient because the fluxon will not pass
 the qubit with $\sigma=1$.
The fluxon delay time (the difference between the times necessary to pass
the $\sigma=1$ and $\sigma=-1$ qubits) depends on the 
impurity length $\rho$, its amplitude $\mu$ and is almost independent on 
the impurity barrier
width $l$. There is an important difference between the
parameters $l$ and $\rho$. The parameter $l$ also enters as the $1/l^2$
prefactor before the amplitude $\mu$ in the equations of motion. Therefore,
it can be treated as a measure of the deviation from the
$\delta'(x)$ limit. From the obtained results we conclude that 
the piece-wise approximation of the $\delta'(x)$ impurity works well. 
On the contrary, the 
parameter $\rho$ influences only the length of the impurity and not
its amplitude.

As far as the future research is concerned, we believe that 
consideration of the two-dimensional JJ is important alongside
with the studies of the Josephson plasmon radiation due to fluxon
scattering on such a finite-size dipole impurity for both the 1D and
2D Josephson junctions.

\section*{Acknowledgemets}
The authors acknowledge the support of the National Academy of 
Sciences of Ukraine through the program No. 0117U000236.

\appendix
\section{Correction calculation}
\label{app1}
In this Appendix the details on how the second order correction 
for $\gamma_{thr}$ is calculated. 

\paragraph{Case $\sigma=1$ ($\mu>0$)}
\par
Taking to account that 
$\text{sech}(-\pi\gamma/\mu)=1-\pi^2\gamma^2/2\mu^2$ and Eq. (\ref{diss})
we can rewrite the energy balance equation:
\begin{equation}\label{a1}
\left(\frac{\pi\gamma}{2\alpha}\right)^2+8\alpha\sqrt{\frac{\mu}{2}}
\left [2 \ln \left(\frac{1+\sqrt{2}}{\sqrt{2}}\right)-
\frac{\pi\gamma}{\mu}\right]
= 2\mu\left [1-\frac{1}{2}\left (\frac{\pi\gamma}{\mu}
\right)^2\right]~.
\end{equation}
We introduce the following small parameters:
\begin{equation} \label{a2}
G=\frac{\pi\gamma}{2|\mu|}\ll 1,
~~\Upsilon=\alpha\sqrt{\frac{2}{|\mu|}}\ll 1~.
\end{equation}
As a result, we get a nonlinear algebraic equation
\begin{equation}\label{a3}
\left(\frac{G}{\Upsilon}\right)^2=-4\Upsilon \left [
\ln\left(\frac{1+\sqrt{2}}{\sqrt{2}}\right)-G\right]+1,
\end{equation}
where the terms of the order ${\cal O}(\Upsilon^2)$,${\cal O}(G^2)$
and higher have been neglected in the right hand side.
It is easy to see that in the lowest order the following
equality holds: $\Upsilon=G$.
Thus, if we want to obtain the second order correction we
look for the small correction to the above formula: 
$G=\Upsilon+\Delta\Upsilon$, $\Delta \Upsilon \ll \Upsilon, G$.
Substitution of this expansion in Eq. (\ref{a3}) yields
\begin{equation}\label{a4}
\Delta \Upsilon+2 \ln \left(\frac{1+\sqrt{2}}{\sqrt{2}}\right)\Upsilon^2=0,
\end{equation}
where the terms ${\cal O}(\Upsilon^3)$ and higher have been neglected.
As a result the final expression for the threshold current reads:
\begin{equation}
G=\Upsilon \left [1-\Upsilon\ln \left(\frac{1+\sqrt{2}}{\sqrt{2}}\right)
 \right],
\end{equation}
or
\begin{equation}
\gamma_{thr}=\frac{2\alpha}{\pi}\left[\sqrt{2\mu}-
4\alpha \ln \left(\frac{1+\sqrt{2}}{\sqrt{2}}\right)\right]~.
\end{equation}

\paragraph{Case $\sigma=-1 (\mu<0)$}
The energy balance equation [see Eq. (\ref{27})] reads:
\begin{equation}\label{a7}
\left(\frac{\pi\gamma}{2\alpha}\right)^2=
8 \left \{\alpha\sqrt{\mu}~
{}_2F_1\left [{1\over 4},{1\over 2},{5\over 4},-\left(\frac{\pi\gamma}{2\mu}\right) ^2\right ]+\frac{\pi\gamma}{2}
\left (1+\ln \frac{\pi\gamma}{2\mu}\right ) \right \}-2\pi\gamma~.
\end{equation}
We introduce parameters $G$ and $\Upsilon$ in the same way as in
Eqs. (\ref{a2}):
\begin{equation}\label{a9}
\left(\frac{G}{\Upsilon}\right)^2=4\sqrt{2}\frac{\Upsilon}{\sqrt{G}}
~{}_2F_1\left ({1\over 4},{1\over 2},{5\over 4},-\frac{1}{G^2}\right)
+4G\ln {G}~.
\end{equation}
With the help of the transformation formulae for the hypergeometric
function \citep{as84} we can rewrite the function with its fourth 
variable becoming $-G^2$. After that it can be expanded in
the Taylor series up to the term ${\cal O}(G^2)$:
\begin{eqnarray}
&&{}_2F_1\left ({1\over 4},{1\over 2},{5\over 4},-\frac{1}{G^2}\right)=
\frac{1}{4}\Gamma^2\left(\frac{1}{4}\right)\left [
\sqrt{\frac{G}{\pi}}{}_2F_1\left({1\over 4},0,{3\over 4},-G^2 \right) 
+\right .\\
\nonumber
&&\left. 
+\frac{1}{\sqrt{2}\pi}{}_2F_1\left({1\over 2},{1\over 4},{5\over 4},-G^2 \right) \right ]
\simeq \frac{1}{4}\Gamma^2\left(\frac{1}{4}\right) \left (
\sqrt{\frac{G}{\pi}}+\frac{G}{\sqrt{2}\pi}\right )~,
\end{eqnarray}
where $\Gamma(x)$ is the gamma function. Taking into account that 
both $G$ and $\Upsilon$ are small parameters,
we can rewrite Eq. (\ref{a9}) as
\begin{equation}\label{a10}
G\simeq G_0+G_1= \left(\frac{2}{\pi}\right )^{1/4}\Gamma \left 
(\frac{1}{4} \right)\Upsilon^{3/2}+G_1,~~ G_1 \ll G_0~.
\end{equation}
In the zero approximation, $G\simeq G_0$ we obtain 
\begin{equation}
\gamma_{thr}^{(0)}=\frac{4}{\pi^{5/4}}\Gamma \left 
(\frac{1}{4} \right)|\mu|^{1/4}\alpha^{3/2}~.
\end{equation}
Substitution of the expansion (\ref{a10}) into (\ref{a9}) yields the 
correction term $G_1$, 
\begin{equation}
G_1=2\Upsilon^2 \ln \left[ \left(\frac{2}{\pi} \right)^{1/4}\Gamma \left 
(\frac{1}{4} \right) \Upsilon^{3/2}\right ]~.
\end{equation}
As a result, we get the full expression for the
threshold current:
\begin{equation}
\gamma_{thr}=\frac{4}{\pi^{5/4}}\Gamma \left 
(\frac{1}{4} \right)|\mu|^{1/4}\alpha^{3/2}+12\frac{\alpha^2}{\pi}
\ln \left[ \left(\frac{2\Gamma\left 
(\frac{1}{4} \right)}{\pi^{1/4}}\right)^{2/3}
\frac{\alpha}{|\mu|^{1/2}}\right]~.
\end{equation}

%
%

\end{document}